\documentclass[a4paper,12pt,onecolumn]{article}
\usepackage[a4paper,top=3cm,bottom=3cm,left=2.0cm,right=2.0cm]{geometry}
\usepackage[utf8]{inputenc}
\usepackage[english]{babel}
\usepackage[T1]{fontenc}
\usepackage{amsmath}
\usepackage[autostyle]{csquotes}
\usepackage{amssymb}
\usepackage{amsfonts,mathrsfs}
\usepackage{amsthm}
\usepackage{graphicx}
\usepackage{dcolumn}
\usepackage{bm}
\usepackage{tabularx}
\usepackage{array}
\usepackage{float}
\usepackage{braket}
\usepackage[usenames,dvipsnames]{xcolor}
\usepackage{afterpage}
\usepackage[style=numeric-comp, sorting=none]{biblatex}
\usepackage{hyperref}

\definecolor{mygreen}{RGB}{3, 150, 0}

\newcommand{\I}{\mathcal{I}}

\newcommand{\C}{\mathcal{C}}

\renewcommand{\P}{\mathcal{P}}
\newcommand{\D}{\mathcal{D}}
\renewcommand{\a}{\hat{a}}
\newcommand{\ad}{\hat{a}^\dagger}
 
\newtheorem{theorem}{Theorem}[section]
\theoremstyle{definition}
\newtheorem{definition}[theorem]{Definition}

\newcommand{\ketbra}[2]{\Ket{#1}\Bra{#2}}
\newcommand{\proj}[1]{\Ket{#1}\Bra{#1}}

\title{\bf Hierarchy of continuous-variable quantum resource theories}
\author{ \normalsize Giulio Gianfelici ,\, Hermann Kampermann, and Dagmar Bru\ss \\ \scriptsize Institut f\"{u}r Theoretische Physik III, Heinrich-Heine-Universit\"{a}t D\"{u}sseldorf, D-40225 D\"{u}sseldorf, Germany }
\date{
\footnotesize \textbf{Email:} \href{mailto:giulio.gianfelici@uni-duesseldorf.de}{giulio.gianfelici@uni-duesseldorf.de}}

\begin{document}
\maketitle 

\begin{abstract}
\noindent Connections between the resource theories of coherence and purity (or non-uniformity) are well known for discrete-variable, finite-dimensional, quantum systems.
We establish analogous results for continuous-variable systems, in particular Gaussian systems.
To this end, we define the concept of maximal coherence at fixed energy, which is achievable with energy-preserving unitaries. We show that the maximal Gaussian coherence (where states and operations are required to be Gaussian) can be quantified analytically by the relative entropy. 
We then propose a resource theory of non-uniformity, by considering the purity of a quantum state at fixed energy as resource, and by defining non-uniformity monotones. In the Gaussian case, we prove the equality of Gaussian non-uniformity and maximal Gaussian coherence. Finally, we show a hierarchy for non-uniformity, coherence, discord and entanglement in continuous-variable systems.
\end{abstract}

\section{Introduction}
Quantum resource theories \cite{chitambar2019quantum} describe the resources of quantum states in a quantitative way. The set of states is divided into \emph{free states} (having no resource) and \emph{resource states}. Quantum operations are called \emph{free} when they transform any free state into a free state, i.e. free operations cannot increase the resources.

Different resource theories have different sets of free states and operations. For instance, the resource theory of coherence \cite{JA06, BCP14, PhysRevA.94.052324, RevModPhys.89.041003, ZSLF16, JX16} identifies states that are diagonal in a certain basis as free, while the resource theory of purity \cite{PhysRevA.67.062104, GMNSH15, SKWGB18} considers the maximally mixed state as the only free state. 
The resourcefulness of a quantum state can be quantified by a resource monotone, which is a function that is non-increasing under free operations. In particular, relevant monotones for several resource theories are based on the relative entropy \cite{JA06, GMNSH15, PhysRevA.78.060303, PhysRevA.57.1619}. 

For discrete-variable (DV) systems, an important connection between coherence, purity, entanglement and discord was found in \cite{SKWGB18}. Here, it was shown that the purity of a quantum state is$\,$---$\,$for an appropriate resource monotone$\,$---$\,$equal to the maximal coherence that can be obtained by applying unitary operations to the state. This quantity then upper-bounds the maximal quantum discord and entanglement of the state. 

This result cannot be straightforwardly extended to continuous-variable (CV) systems. Infinite-dimensional Hilbert spaces are structurally different from their finite-dimensional counterpart \cite{olivares2012quantum, WPGC12, adesso2014continuous, serafini2017quantum}, and this difference influences the mathematical definition of the physical quantities themselves and their resource theories \cite{lami2018gaussian}. In particular, infinite-dimensional Hilbert spaces allow the generation of infinite resources via unitary operations \cite{ESP02, keyl2002infinitely}. 

A common strategy to replicate DV results is to introduce valid physically and experimentally motivated constraints. The first restriction is to consider finite energy \cite{ESP02, AE05, AE11} and to explore the use of energy-preserving unitaries, as those operations are easily available in laboratories. The second restriction is to focus on Gaussian states and operations, as most of the relevant phenomena in quantum information and quantum optics can be described by at most quadratic Hamiltonians. 

Equipped with these assumptions, we investigate the resource theories of CV coherence \cite{ZSLF16, JX16}. We discuss both general and Gaussian coherence, the latter by restricting the set of quantum states and operations to be Gaussian. We define the concept of maximal coherence of a quantum state at fixed energy, as the coherence that can be obtained via applying energy-preserving unitaries. In the case of Gaussian coherence, we find the structure of the states with maximal coherence and discuss the form of the maximizing unitary in some specific cases. We derive an analytical expression for the relative entropy of the maximal Gaussian coherence. We then propose a resource theory of non-uniformity, to describe purity at fixed energy as resource, considering states that maximise the entropy at fixed energy as free. However, our resource theory studies the interactions of a quantum system with a noisy thermal environment, therefore it is connected to the resource theories of a-thermality \cite{SLLSHA20,NABTYG19}, where states out of thermal equilibrium are identified as resources. Our theory emphasises the entropic exchanges between the system and the environment, rather than the energetic ones. 

Finally, we establish a connection between non-uniformity, coherence, quantum discord and entanglement, by identifying a hierarchy between them. In particular, the maximal Gaussian coherence is bounded by the Gaussian non-uniformity and both upper-bound the maximal discord and the entanglement. Our results represent an extension to infinite dimensions of the hierarchy found in \cite{SKWGB18}.

We begin our work by introducing our setting together with the basic notions of CV quantum information in Sec. \ref{sec:prel}. We review coherence and Gaussian coherence in Sec. \ref{sec:Gaucoh}. In Sec. \ref{sec:MCMS}, we define the maximally coherent mixed state at fixed energy and derive its properties for the Gaussian case. We assemble the resource theory of non-uniformity in Sec. \ref{sec:GNU} and illustrate the connections between non-uniformity, coherence, discord and entanglement in Sec. \ref{sec:conn}. Finally, we summarise our results in Sec. \ref{sec:conc}.
 
\section{Notation and preliminaries}
\label{sec:prel}
In the following, we will indicate vectors and matrices as bold lowercase and uppercase letters, respectively. We shall consider systems with a finite number of discrete \emph{spectral} and \emph{spatial} modes, which refer to the frequency and location of the mode, respectively. We label the mode operators of a mode with two indices: an index $\omega$ for the spectral degrees of freedom and an index $j$ for the spatial degrees of freedom. 

\begin{figure}[htb]
  \centering
   \includegraphics[scale=0.3]{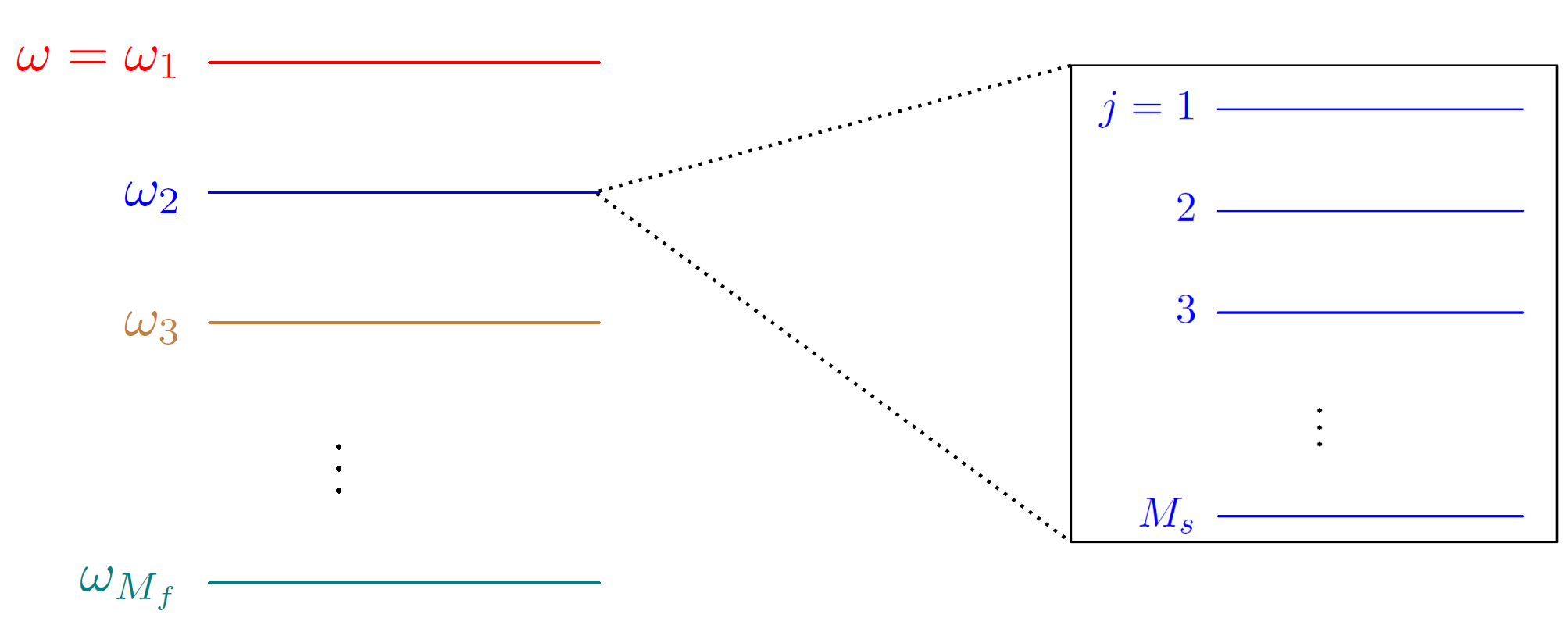}
  \caption{Graphical depiction of the labeling of the spectral and spatial modes. The modes are cataloged first in terms of their frequency $\omega=\omega_1, \omega_2, \dots \omega_{M_f}$ (represented by a distinct colour) and then in terms of their spatial label $j=1,2,\dots M_s$.}
    \label{fig:modes}
\end{figure}

We sort our mode operators by gathering all mode operators with the same frequency, i.e. 
\begin{equation}
\label{eq:freqmode}
\set{\hat{\bm{a}}_{\omega_1},\hat{\bm{a}}_{\omega_2}, \dots \hat{\bm{a}}_{\omega_{M_f}}}, 
\end{equation}
where $M_f$ is the number of different frequencies, and for each frequency $\omega$
\begin{equation}
\label{eq:spatmode}
\hat{\bm{a}}_{\omega}= (\a_{\omega; 1},\a_{\omega; 2},\dots, \a_{\omega; M_s})^T, 
\end{equation}
with $\a_{\omega; j}$ being the annihilation operator for a mode with spectral label $\omega$ and spatial label $j$, and $M_s$ being the total number of spatial labels. Without loss of generality, we assume that $M_s$ is the same for all frequencies. A graphical depiction of our mode labeling is drawn in Fig. \ref{fig:modes}.

The operators satify the usual bosonic commutation relations
\begin{align}
&\Big[\a_{\omega; j}, \a_{\omega'; j'}\Big]=\Big[\a^\dagger_{\omega; j}, \a^\dagger_{\omega'; j'}\Big]= 0,\label{eq:CCR1}\\
 &\Big[\a_{\omega; j}, \a^\dagger_{\omega'; j'}\Big]= \delta_{\omega\omega'} \delta_{jj'}.\label{eq:CCR2}
\end{align}

Using the notation of Eqs. \eqref{eq:freqmode} and \eqref{eq:spatmode}, the free Hamiltonian of the system reads 
\begin{equation}
\label{eq:freeHam}
 \hat{H}=\sum_{\omega=\omega_1}^{\omega_{M_f}}\,\hbar\omega\:\left[ \sum_{j=1}^{M_s}\left(\hat{a}_{\omega;j}^\dagger\, \hat{a}_{\omega;j}+\frac{1}{2}\right)\right]. 
\end{equation}
In the following, we will always assume that the system has finite mean energy, i.e. that the Hamiltonian satisfies $\braket{\hat{H}}<\infty$. This is a natural and physically reasonable assumption, and a mathematically necessary precondition for the trace-norm continuity of many functionals \cite{ESP02, AE05, AE11}.

The total number of modes is $M=M_f M_s$. We shall label the modes with the index $m \equiv \omega; j$ whenever there is no need to distinguish between spectral and spatial modes. We use the order $m=(\omega_1;1),(\omega_1;2),\dots, (\omega_1;M_s), (\omega_2;1),\dots, (\omega_{M_f};M_s)$.

We now recall basic concepts of Gaussian quantum information, inspired from \cite{olivares2012quantum, WPGC12, adesso2014continuous,serafini2017quantum} and written according to our notation. 

For each mode $m$, we define the canonical conjugate operators $\hat{q}_m:=(\a_m +\ad_m)/\sqrt{2} $ and $ \hat{p}_m:= (\a_m -\ad_m)/(i\sqrt{2})$. We group them in a vector $\hat{\bm{r}}:=(\hat{q}_1,\hat{p}_1,\hat{q}_2,\hat{p}_2,\dots,\hat{q}_M,\hat{p}_M)^T$.

The first and second moment of a state $\rho$ is the \emph{displacement vector} $\bm{d}$ and the \emph{covariance matrix} $\bm{V}$, respectively. Their components $d_m$ and $V_{mm'}$ read in terms of components $\hat{r}_m$ of $\hat{\bm{r}}$ as \cite{adesso2014continuous}: 
\begin{align}
\label{eq:disvec}
    &d_{m}:= \braket{\hat{r}_m}=\mathrm{Tr}\,[\hat{r}_m\,\rho], \\
\label{eq:covmat}
&V_{m m'}:=\,\braket{\hat{r}_m\,\hat{r}_{m'}+\hat{r}_{m'}\,\hat{r}_m}-2\,d_m\,d_{m'}.
\end{align}

Both moments are real, and $\bm{V}$ is symmetric and positive definite. Gaussian states are represented by a Gaussian quasi-probability distribution in the phase space and are fully characterised by the first and second moments. The $2M$-dimensional displacement vector and the $2M\times 2M$ covariance matrix of any Gaussian state can be written in the following block form:
\begin{equation}
\label{eq:genmat}
\bm{d}=
\begin{pmatrix}
 \bm{d}_1 \\
 \bm{d}_2 \\
 \vdots \\
 \bm{d}_M
\end{pmatrix}
,\qquad
 \bm{V}=
 \begin{pmatrix}
  \bm{V}_1 & \bm{\Delta}_{12} & \dots & \bm{\Delta}_{1M} \\
  \bm{\Delta}^T_{12} & \bm{V}_2 & \dots & \bm{\Delta}_{2M} \\
  \vdots & \vdots &\ddots & \vdots \\
  \bm{\Delta}^T_{1M} & \bm{\Delta}^T_{2M} & \dots & \bm{V}_{M}
 \end{pmatrix},
\end{equation}
where $\bm{d}_m$ are $2$-dimensional vectors, and $\bm{V}_m$ and $\bm{\Delta}_{m m'}$ are $2\times 2$ real matrices. In particular, $\bm{d}_m$ ($\bm{V}_m$) corresponds to the displacement vector (the covariance matrix) of the reduced state  $\rho_m=\mathrm{Tr}_{\bm{m}\backslash m}\left[\,\rho\,\right]$ after partial trace of all modes but the $m$-th, while $\bm{\Delta}_{m m'}$ is related to the correlations between the modes $m$ and $m'$  \cite{olivares2012quantum}.

The total average occupation number can be derived as 
\begin{equation}
\label{eq:ni}
\braket{\hat{N}}=\sum_{m=1}^M\,\bar{n}_m=\sum_{m=1}^M\,\frac{1}{4}\,(\mathrm{Tr}[\bm{V}_m]+ 2|\bm{d}_m|^2-2),
\end{equation} 
where $\bar{n}_m$ is the average occupation number of the $m$-th mode. This expression can be obtained by writing $\mathrm{Tr}[\bm{V}_m]$ in terms of Eq. \eqref{eq:covmat} and the mode operators $\a_m$ and $\ad_m$
\begin{equation}
 \begin{split}
  \mathrm{Tr}[\bm{V}_m]&=2\braket{\hat{q}_m^2}+2\braket{\hat{p}_m^2}-2\braket{\hat{q}_m}^2-2\braket{\hat{p}_m}^2 = \braket{(\a_m+\ad_m)^2}-\braket{(\a_m-\ad_m)^2}-2|\bm{d}_m|^2 \\
  &= 2+4\braket{\ad_m\a_m}-2|\bm{d}_m|^2.
 \end{split}
\end{equation}
Eq. \eqref{eq:ni} follows by setting $\braket{\hat{N}}=\sum_{m=1}^M\,\braket{\ad_m\a_m}$.
Notice that this formula holds regardless of the presence of correlations between different modes, for both Gaussian and non-Gaussian states. 

By Williamson's theorem \cite{JW36}, any covariance matrix $\bm{V}$ can be brought into a diagonal form:
\begin{equation}
\label{eq:WT}
 \bm{V}=\bm{S} \,\bm{D}\,\bm{S}^T,
\end{equation}
where $\bm{D}$ is a diagonal matrix 
\begin{equation}
 \bm{D}= \text{diag}\left[\nu_1, \nu_1, \dots, \nu_M,\nu_M\right],
\end{equation}
and $\bm{S}$ is a \emph{symplectic matrix}, i.e. a real matrix that satisfies
\begin{equation}
 {\bm S}{\bm \Omega} {\bm S}^T = \bm{\Omega}, \qquad
  \bm{\Omega}=\bigoplus_{m=1}^M\,
 \begin{pmatrix}
  0 & 1 \\
  -1 & 0
 \end{pmatrix}.\label{eq:symp}
\end{equation}
The variables $\nu_m\geq 1$ are called \emph{symplectic eigenvalues} and obey the Bose-Einstein statistics: 
 \begin{equation}
 \label{eq:BES}
  \nu_m=\nu_{\omega;j}=\frac{1}{2}\,\left[\exp\left(\frac{\hbar\omega}{\kappa\,T_{\omega;j}}\right)-1\right]^{-1}-\frac{1}{2},
 \end{equation}
 where $\kappa$ is the Boltzmann constant and $T_{\omega;j}$ is the temperature of the $m$-th mode, with $m=(\omega;j)$. The symplectic eigenvalues are used to express several properties of Gaussian states. For instance, the von-Neumann entropy of a Gaussian state reads \cite{HSH99}:
\begin{equation}
\label{eq:entr}
S(\rho)=\sum_{m=1}^M\,\left(\frac{\nu_m+1}{2}\,\log\frac{\nu_m+1}{2}-\frac{\nu_m-1}{2}\,\log\frac{\nu_m-1}{2}\right).
\end{equation}

Any unitary that preserves the Gaussianity of quantum states is called \emph{Gaussian unitary}. In terms of the moments, a Gaussian unitary acts as \cite{serafini2017quantum}
\begin{equation}
\label{eq:Gausunit} 
 \begin{split}
 \bm{d}&\rightarrow \bm{S}\,\cdot {\bm d}+\bm{v}, \\ 
 \bm{V}&\rightarrow \bm{S}\cdot\bm{V}\cdot\bm{S}^T, 
 \end{split}
 \end{equation}
where $\bm{v}$ is a $2M$-dimensional vector and $\bm{S}$ is a $2M\times 2M$ symplectic matrix.

\begin{theorem}[\emph{Bloch-Messiah decomposition} \cite{DMS95}]
 Any $2M\times 2M$ symplectic matrix can be decomposed as
 \begin{equation}
 \label{eq:BloMes}
 \bm{S}=\bm{O}_1\,\left[\bigoplus_{m=1}^M\, \bm{Z}(r_m) \right]\,\bm{O}_2,
 \end{equation}
 where the $2M\times 2M$ matrices $\bm{O}_1$ and $\bm{O}_2$ are symplectic and orthogonal (we generally denote symplectic orthogonal matrices as $\bm{O}$), and the $2 \times 2$ matrix $\bm{Z}(r_m)$ is a single-mode squeezer with squeezing parameter $r_m$, that is
 \begin{equation}
 \label{eq:squeez}
  \bm{Z}(r_m):= 
  \begin{pmatrix}
   e^{-r_m} & 0 \\
   0 & e^{r_m}
  \end{pmatrix}
  .
 \end{equation}
\end{theorem}

For $r_m=0$ (absence of squeezing), $\bm{Z}(r_m)$ becomes the identity $\bm{I}$. Therefore, in Eq. \eqref{eq:BloMes}, $\bm{S}=\bm{O}_1\bm{O}_2$ is orthogonal, being the product of two orthogonal matrices.
\begin{definition}
\label{def:pasGau}
A \emph{passive unitary} \cite{WPGC12,olivares2012quantum} is a Gaussian unitary $\hat{U}_{\bm{O}}$ that is represented in the phase space (Eq. \eqref{eq:Gausunit}) by $\bm{v}=\bm{0}$ and a symplectic orthogonal matrix $\bm{O}$ in the form of
 \begin{equation}
\label{eq:blockpas}
 \bm{O}=\bigoplus_{\omega=\omega_1}^{\omega_{M_f}}\,\bm{O}_\omega,
\end{equation}
where $\bm{O}_\omega$ are $2 M_s\times 2 M_s$ symplectic orthogonal matrices acting on the subset of modes with frequency $\omega$. Conversely, any Gaussian unitary that is not passive, is called \emph{active}.
\end{definition}

Active unitaries are associated with linear displacements and squeezing. Passive unitaries are realised by linear-optics circuits, that is any multiport interferometer made of beam splitters and phase shifters. They preserve the total average occupation number (see Eq. \eqref{eq:ni}):
\begin{equation}
\braket{\hat{N}}=\frac{1}{4}\,(\mathrm{Tr}[\bm{O}\,\bm{V}\,\bm{O}^T]+ 2|\bm{O} \bm{d}|^2-2M)=
\frac{1}{4}\,(\mathrm{Tr}[\bm{V}]+ 2| \bm{d}|^2-2M).
\end{equation}
Since $\hat{N}$ commutes with the Hamiltonian $\hat{H}$ of the system, passive Gaussian unitaries preserve the average energy, too. Passive Gaussian unitaries are the only energy-preserving Gaussian unitaries \cite{adesso2014continuous, serafini2017quantum}. They create correlations between spatial modes with the same frequency $\omega$. They do not allow interactions between modes with different frequencies (see Eq. \eqref{eq:blockpas}, and \cite{fabre2020modes}).

Gaussian unitaries are not the only operations that preserve the Gaussianity of a quantum state. A \emph{Gaussian channel} is a completely positive trace-preserving (CPTP) operation that maps Gaussian states into Gaussian states. In terms of the moments, a Gaussian channel acts as \cite{holevo2001evaluating}
\begin{equation}
\label{eq:Gauschan}
 \begin{split}
  \bm{d}&\rightarrow \bm{T}\,\cdot {\bm d}+\bm{v}, \\
  \bm{V}&\rightarrow \bm{T}\cdot\bm{V}\cdot\bm{T}^T+\bm{N},
 \end{split}
\end{equation}
where $\bm{T}$, $\bm{N}$ are $2M \times 2M$ real matrices, $\bm{N}\geq 0$, and $\bm{v}$ is a $2M$-dimensional vector.

\section{Resource theory of (Gaussian) coherence}
\label{sec:Gaucoh}

A state $\rho$ of $M$ bosonic modes is said to be \textit{incoherent} if it is diagonal in the M-mode Fock basis \cite{ZSLF16}, i.e.
\begin{equation}
\label{eq:geninch}
\rho=\sum_{n_1\dots n_M=0}^\infty\,p_{n_1\dots n_M} \ketbra{n_1}{n_1}\otimes\dots\otimes\ketbra{n_M}{n_M},
\end{equation}
for an arbitrary set of non-negative probabilities $\set{p_{n_1\dots n_M}}$.

We denote the set of all incoherent states by $\I$. The resource theory of coherence admits different sets of free operations. The maximal set of free operations for $\I$ are called \emph{maximally incoherent operations} (MIO) \cite{JA06}. These are all maps that cannot create coherence, i.e., 
\begin{equation}
\label{eq:MIO}
 \Lambda_{MIO}(\rho)\in\I,\qquad \forall\:\rho\in\I.
\end{equation}

An extensive study of MIO in infinite-dimensional Hilbert spaces has not been carried out so far. We shall not investigate this here, referring to \cite{RevModPhys.89.041003} for a general review of coherence and to \cite{ZSLF16} for the CV case.

\begin{definition}
\label{def:cohmes}
For continuous-variable systems, a function $\C(\rho)$ is a suitable measure of coherence with respect to a chosen set of free operations, e.g. MIO (see Eq.\eqref{eq:MIO}), if it satisfies the following properties \cite{BCP14,ZSLF16,JX16}: 
 \begin{description}
  \item[(C1)] Positivity: $\C(\rho)\geq 0$ for any density operator $\rho$ and $\C(\rho)= 0$ iff $\rho\in\I$;
  \item[(C2)] Monotonicity under the chosen set of free operations, e.g. for MIO: 
\begin{equation}
\C(\rho)\geq \C(\Lambda_{MIO}(\rho));
\end{equation}
  \item[(C3)] Convexity: 
  \begin{equation}
   \sum_n\,p_n\,\C(\rho_n)\geq \C\left(\sum_n\,p_n\,\rho_n\right);
  \end{equation}
  \item[(C4)] Finite coherence for systems with finite energy: 
  \begin{equation}
\braket{\hat{H}}<\infty\Rightarrow\C(\rho)<\infty.                                                 
  \end{equation} 
 \end{description}
\end{definition}
The condition (C4) is specific for CV systems. Denoting with $S(\rho\|\tau)$ the quantum relative entropy between $\rho$ and $\tau$,
\begin{equation}
 S(\rho\|\tau):={\rm Tr}[\rho\log\rho]-{\rm Tr}[\rho\log\tau],
\end{equation}
the \emph{relative entropy of coherence} is defined as:
\begin{equation}
\label{eq:Crho}
 \C_{rel}(\rho):=\min_{\tau\in\I}\,S(\rho\|\tau).
\end{equation}
This measure satisfies all conditions of Def. \ref{def:cohmes} \cite{ZSLF16}.  

The generic CV coherence has not been deeply studied, mainly because of theoretical and experimental difficulties associated with general bosonic Hilbert spaces. We shall therefore address now the relevant Gaussian subclass, in which all states and operations are Gaussian \footnote{In principle, one could consider mixed scenarios, e.g. with Gaussian states and general operations, or vice versa. This goes beyond the scope of our work.}. 

In the realm of Gaussian states, a state is diagonal in the Fock basis if and only if it is a thermal state \cite{JX16}, defined as  
\begin{equation}
 \begin{split}
  &\tau_M(\bar{\bm n}):=\bigotimes_{m=1}^M \tau(\bar{n}_m);\\
  &\tau(\bar{n}_m):=\sum_{n_m=0}^\infty\,\frac{\bar{n}_m^{n_m}}{(\bar{n}_m+1)^{n_m+1}}\,\proj{n_m},
 \end{split}
 \label{eq:therdef}
 \end{equation}
where $\bar{\bm n}=(\bar{n}_1,\dots,\bar{n}_M)$ and $\bar{n}_m$ is the average occupation number of the $m$-th mode (see Eq. \eqref{eq:ni}). The subscript $M$ in $\tau_M$ denotes the number of modes of $\tau_M$, and is omitted for single-mode thermal states. Thermal states have a zero displacement vector and a diagonal covariance matrix 
\begin{equation}
\label{eq:therCM}
\bm{V}\left[\tau_M(\bar{\bm n})\right]=\bigoplus_{m=1}^M\,(2\bar{n}_m+1)\bm{I}.
\end{equation}

We denote the subset of all incoherent Gaussian states by $\I_G$. Xu \cite{JX16} introduced \emph{incoherent Gaussian operations} (IG). 
\begin{definition}
\label{def:IGO}
Incoherent Gaussian operations (IG) are defined as all Gaussian channels $\Lambda_{IG}$ that map thermal states (Eq. \eqref{eq:therdef}) into thermal states, i.e. the maximal set of free operations in this scenario. A generic IG can be written in the form of Eq. \eqref{eq:Gauschan}, where
 \begin{itemize}
 \item $\bm{v}_{IG}=0$;
 \item $\bm{N}_{IG}$ is diagonal, 
 \begin{equation}
  \label{eq:IGON}
  \bm{N}_{IG}=\text{diag}\,\{w_1\,\bm{I}_2, \dots, w_M\,\bm{I}_2\},
 \end{equation}
with $\omega_m\geq 0$;
 \item $\bm{T}_{IG}$ is composed of $M_f$ submatrices $\bm{T}_\omega$ that act on single frequency sectors, namely:
 \begin{equation}
  \bm{T}_{IG}=\bigoplus_{\omega=\omega_1}^{\omega_{M_f}}\,\bm{T}_\omega,
 \end{equation}
 and each $\bm{T}_\omega$ can be generated as follows: 
 \begin{enumerate}
  \item Take $M_s$ real coefficients $t_{\omega; j}\in\mathbb{R}$;
  \item Take $M_s$ $2\times 2$ orthogonal matrices $\bm{\mathcal{O}}_{\omega; j}$, which do not need to be symplectic;
   \item $\bm{T}_\omega$ is given by a permutation of the columns of $\bigoplus_{j=1}^{M_s} t_{\omega; j}\bm{\mathcal{O}}_{\omega; j}$.
 \end{enumerate}
\end{itemize}

\end{definition}

\begin{definition}
 \label{eq:Gauscohmeas}
A function $\C^G(\rho)$ is a suitable measure of Gaussian coherence with respect to IG ($\I_G$) as free operations (free states), if it satisfies the properties (C1)-(C4) of Def. \ref{def:cohmes}. Here, IG are defined in Def. \ref{def:IGO} and states in $\I_G$ are defined by Eq. \eqref{eq:therdef}. This function quantifies the Gaussian coherence, which is, in general, an upper bound for the general coherence (since $\I_G \subset \I$).
\end{definition}

In particular, the \emph{relative entropy of Gaussian coherence} $\C_{rel}^G(\rho)$ can be defined as the relative entropy of coherence (see Eq. \eqref{eq:Crho}) by performing the minimization over $\I_G$. For $M$-mode Gaussian systems, it reads \cite{JX16}:  
\begin{equation}
\label{eq:CSrho}
\begin{split}
 \C_{rel}^G(\rho)&:=S(\rho\|\tau_M(\bar{\bm n}_\rho))= -S(\rho)+\sum_{m=1}^M\, \left[(\bar{n}_m+1)\log (\bar{n}_m+1) -\bar{n}_m\log \bar{n}_m\right],
 \end{split}
\end{equation}
where $\bar{n}_m$ is the reduced average occupation number of $\rho$ (see Eq. \eqref{eq:ni}), $\tau_M(\bar{\bm n}_\rho)$ represents the thermal state with the same $\bar{n}_m$ of $\rho$ and the von-Neumann entropy $S(\rho)$ is given by Eq. \eqref{eq:entr}. 

\section{Maximally coherent mixed states at fixed energy}
\label{sec:MCMS}
Coherence is a basis-dependent quantity, therefore it is affected by unitary operations. For a given DV state $\rho$, the \emph{maximally coherent mixed state} (MCMS) \cite{SBDP15, YDGLS16} is defined as $\rho_{max}=V\,\rho\,V^\dagger$, where $V$ is the unitary that maximises the coherence of $\rho$.

For CV systems, this definition is not applicable, since the coherence depends on the energy of the system. This can be seen, for instance, in Eq. \eqref{eq:CSrho} for the relative entropy of Gaussian coherence. Therefore, energy non-preserving unitaries can in principle increase the coherence indefinitely. 

From an experimental point of view, energy-preserving unitaries are easier to realise and do not require interaction with an external source of energy.

This is the motivation to define a family of \emph{maximally coherent mixed states at fixed energy}:
\begin{definition}
A state $\rho_{max}$ is a \emph{maximally coherent mixed state} (MCMS) \emph{at fixed energy} with respect to a coherence monotone $\C$ (see Def. \ref{def:cohmes}) if
\begin{equation}
\label{maxcoh}
 \C(\rho_{max})=\C_{max}(\rho):=\sup_{\hat{U}_{EP}}\, \C(\hat{U}_{EP}\,\rho\, \hat{U}_{EP}^\dagger),
\end{equation}
where $\hat{U}_{EP}$ are energy-preserving unitaries. If we consider Gaussian states and operations in Eq. \eqref{maxcoh}, then $\rho_{max}$ is the \emph{maximally coherent mixed Gaussian state} (MCMGS) \emph{at fixed energy} with respect to $\C$. 
\end{definition}
Let us now focus on the Gaussian case. Passive Gaussian unitaries $\hat{U}_{\bm{O}}$ (see Def. \ref{def:pasGau}) are the only energy-preserving Gaussian unitaries, i.e. they preserve $N_\omega=\sum_j \bar{n}_{\omega; j}$, and thus $N=\sum_\omega\, N_\omega$. However, the interaction between modes with the same frequency $\omega$ results in a redistribution of $\bar{n}_{\omega; j}$. 

Let us consider a generic Gaussian state $\rho$ and call $\rho_\omega=\mathrm{Tr}_{\bm{\omega}\backslash\omega}\,(\rho)$ the state obtained by tracing out all modes with any frequency but $\omega$. From its definition in Eq. \eqref{eq:Crho}, the relative entropy of Gaussian coherence of $\rho$ can be written as the sum of the relative entropies for all $\rho_\omega$:
\begin{equation}
 \C_{rel}^G(\rho)=\sum_{\omega=\omega_1}^{\omega_{M_f}}\C_{rel}^G(\rho_\omega).
\end{equation}
Therefore, the maximal Gaussian coherence can be obtained by maximizing the Gaussian coherence for each $\rho_\omega$. 

\begin{theorem}
\label{theo:mcmsS}
 Among all Gaussian states $\rho_\omega$ with a given symplectic spectrum $\Set{\nu_1,\dots,\nu_{M_s}}$ and a given average occupation number $N_\omega=\sum_j\, \bar{n}_{\omega; j}$, the states with equidistributed reduced average occupation numbers, i.e. $\bar{n}_{\omega; j}=N_\omega/M_s$, $\forall\, j$, are the maximally coherent mixed Gaussian states with respect to the relative entropy of Gaussian coherence $\C^G_{rel}(\rho_\omega)$ (see Eq. \eqref{eq:CSrho}).
\end{theorem}

The proof is given in Appendix \ref{proof:mcmsS}, using Lagrange multipliers. Combining this result with Eq. \eqref{eq:CSrho}, it follows that the maximal relative entropy of Gaussian coherence of a Gaussian state $\rho$ reads
\begin{equation}
\label{cohmax}
\begin{split}
 \C^G_{rel;\:max}(\rho):=& S(\rho_{max}\|\tau_M(\bar{\bm n}_{\rho_{max}}))\\
 =&-S(\rho)+\sum_{\omega=\omega_1}^{\omega_{M_f}}\,\left[(N_\omega+ M_s)\log\left(\frac{N_\omega+M_s}{M_s}\right)-N_\omega\log\left(\frac{N_\omega}{M_s}\right)\right],
\end{split}
\end{equation}
where $\tau_M(\bar{\bm n}_{\rho_{max}})$ is the $M$-mode thermal state with occupation numbers $\bar{\bm n}_{\rho_{max}}$ (see Eq. \eqref{eq:therdef}). In Appendix \ref{proof:dist}, we provide an alternative analytical expression for $\C^G_{rel;\:max}$ (see Theorem \ref{theo:dist}), that will play a pivotal role in the next sections.

\begin{theorem}
\label{theo:dist}
The maximal relative entropy of Gaussian coherence $\C^G_{rel;\:max}(\rho)$ of any Gaussian state $\rho$ can be expressed as:
 \begin{equation}
  \C^G_{rel;\:max}(\rho)=\sum_{\omega=\omega_1}^{\omega_{M_f}}\, S\left(\rho_\omega\,\Big\|\,\tau_{M_s}\left(\frac{N_\omega}{M_s},\dots, \frac{N_\omega}{M_s}\right)\right),
 \end{equation}
where $\tau_{M_s}(N_\omega/M_s,\dots,N_\omega/M_s)$ is the thermal state (see Eq. \eqref{eq:therdef}) with $\bar{n}_{\omega; j}=N_\omega/M_s$ for all $j=1,2,\dots M_s$.
\end{theorem}

Finally, in Appendix \ref{proof:examples}, we provide examples of passive unitaries that maximise the Gaussian coherence for two generic classes of Gaussian states.

\section{Resource theory of (Gaussian) non-uniformity}
\label{sec:GNU}

The resource theory of purity (or non-uniformity) \cite{GMNSH15} belongs to a family of resource theories of quantum thermodynamics in which states out of some form of equilibrium are considered as resources \cite{janzing2000thermodynamic, PhysRevLett111250404, lostaglio2019introductory, whitney2018thermodynamics}. Usually, this equilibrium is given by assuming the environment at a certain background temperature $T$: the free states are thermal states at the same temperature and the free operations are those which are generated by an energy-preserving unitary acting on the system and the environment.

For DV systems, the resource theory of purity arises when the Hamiltonian is fully degenerate at any temperature. Then all unitaries become energy preserving (hence free operations) and the exchanges between the system and the environment are purely entropic \cite{GMNSH15}. The state representing informational equilibrium becomes the maximally mixed state (MMS) $\bm{I}/d$, where $d$ is the dimension of the system. 
The MMS is the only free state, as every other state possesses some non-uniformity. 

The DV theory cannot be straightforwardly extended to Gaussian systems, because a proper MMS is nonphysical, as it is associated with infinite energy in infinite-dimensional Hilbert spaces \cite{serafini2017quantum}. 

While several resource theories of Gaussian states out of thermal equilibrium exist \cite{SJVC19, SLLSHA20, NABTYG19}, we choose a different approach, that emphasises the informational aspects of the interactions between the system and the environment over the energetic ones. 
We consider purity at given energy as a resource and we refer to this resource theory as \emph{non-uniformity}: with a similar argument as given by Gour et al \cite{GMNSH15}, we use this term in place of "purity" because we shall consider pure states at different energy as states with different resource content.

Consider an $M$-mode state $\rho$, with $M_f$ frequencies and $M_s$ spatial labels (see Sec. \ref{sec:prel}). Consider also a Hamiltonian $\hat{H}$ in the form of Eq. \eqref{eq:freeHam}. In terms of the frequencies its mean energy can be written as 
\begin{equation}
 \label{eq:Homega}
 \braket{\hat{H}}=\sum_{\omega=\omega_1}^{\omega_{M_f}}\,E_\omega,
\end{equation}
where $E_\omega= \hbar\omega N_\omega$ (see Eq.\eqref{eq:freeHam}) and any other contribution is set to zero by a suitable choice for the zero-point energy. In the DV resource theory of purity, a dimension $d$ for the set of all states was fixed. In our resource theory, we fix a set of frequencies $\omega=\omega_1,\omega_2,\dots,\omega_{M_f}$ and the energy in each frequency mode $E_\omega$, thus fixing $\braket{\hat{H}}$ as in Eq. \eqref{eq:Homega}. The states and modes can have any temperature that is compatible with $E_\omega$, i.e. the thermal component of the energy cannot be higher than $E_\omega$ for any frequency sector. 

For a single frequency $\omega$, the state that maximises the entropy is the natural CV counterpart of the maximally mixed state in the DV case. We call it the \emph{uniform state at frequency $\omega$.} In our setting, with $M_f$ different frequencies, we consider as free state the tensor product of all uniform states at frequency $\omega$ over all frequencies. We call it the \emph{uniform state}. 

From Eq. \eqref{eq:ni}, we see that the average energy $E_\omega=\hbar\omega\,N_\omega$ is a function of the first and second moments of $\rho$, for both Gaussian and non-Gaussian states. It is well known that Gaussian states attain the maximum von-Neumann entropy among all states having the same displacement vector and covariance matrix \cite{HSH99}. Therefore, even if we consider the set of all CV states, we can search for the uniform state in the subset of Gaussian states.

For single-mode Gaussian systems at a fixed energy $E_\omega=\hbar\omega N_{\omega}$, the von-Neumann entropy is maximised by Gaussian thermal states with average occupation number $N_{\omega}$ \cite{braunstein2005quantum}. For $M_s$ spatial modes, we prove in Appendix \ref{proof:unif} the following result: 
\begin{theorem}
\label{theo:unif}
 For a quantum system of $M_s$ spatial modes, the uniform state at frequency $\omega$, i.e. the state that maximises the entropy, is the Gaussian thermal state (see Eq. \eqref{eq:therdef}) with equal single-mode occupation numbers, i.e.
\begin{equation}
 \tau_{M_s}(\bm{\delta}_\omega):=\tau(\delta_\omega)\otimes\tau(\delta_\omega)\otimes\dots\otimes \tau(\delta_\omega),
\end{equation}
where $\bm{\delta}_\omega=(\delta_\omega,\delta_\omega,\dots,\delta_\omega)$,  $\delta_\omega:=N_\omega/M_s$, and $N_\omega$ is the total occupation number for all spatial modes with frequency $\omega$. Considering both spatial and frequency modes, the uniform state becomes 
\begin{equation}
\label{eq:Munifdef}
\tau_M(\bm{\delta})=\bigotimes_{\omega=\omega_1}^{\omega_{M_f}}\,\tau_{M_s}(\bm{\delta}_\omega), 
\end{equation}
 where $M_f$ is the total number of frequencies. The covariance matrix of $\tau_M(\bm{\delta})$ reads
\begin{equation}
\label{eq:unifCM}
 \bm{V}\left[\tau_M(\bm{\delta})\right]=\bigoplus_{\omega=\omega_1}^{\omega_{M_f}}\,\left(2\delta_\omega+1\right)\,\bm{I}_{2M_s}.
\end{equation}
\end{theorem}
This result has an intuitive explanation. The von-Neumann entropy of a Gaussian state (see Eq. \eqref{eq:entr}) depends solely on the symplectic eigenvalues, and the $m$-th eigenvalue is a function of the $m$-th mode's temperature (see Eq. \eqref{eq:BES}). To maximise the entropy, we need to consider thermal states. Among thermal states, the uniform state is defined as the state with the most homogeneous distribution of single-mode energies.

Both the general and the Gaussian version of this resource theory have the same set of free states. We could, in principle, distinguish them via the set of free operations.
\begin{definition}
\label{def:UPO}
The \emph{uniformity-preserving operations} (UP) are all maps that preserve the uniform state $\tau_M(\bm{\delta})$ (see Theorem \ref{theo:unif}), i.e.
\begin{equation}
 \Lambda_{UP}(\tau_M(\bm{\delta}))=\tau_M(\bm{\delta}).
\end{equation} 
We call the Gaussian channels in UP the \emph{uniformity-preserving Gaussian operations} (UPG). 
\end{definition}

Clearly $UPG\subseteq UP$, but we do not know whether this inclusion is strict. Concerning Gaussian operations, a more practical set of free operations is that of \emph{Gaussian noisy operations} (GN), i.e. Gaussian channels $\Lambda_{GN}$ that admit the following decomposition: 
\begin{equation}
\label{eq:GNO}
\Lambda_{GN}[\rho]=\mathrm{Tr}_{M_E}\left[\hat{U}^{(M+M_E)}_{\bm O}\,\left(\rho\otimes\tau_{M_E}(\bm{\delta})\right)\, \hat{U}_{\bm O}^{(M+M_E)\,\dagger}\right],
\end{equation}
where $\tau_{M_E}(\bm{\delta})$ is the uniform state (see Theorem \ref{theo:unif}) for $M_E$ environmental modes and with the same $\bm{\delta}$ of the system (see Eq.\eqref{eq:Munifdef}), and $\hat{U}^{(M+M_E)}_{\bm{O}}$ is an $(M+M_E)$-mode passive Gaussian unitary.

Gaussian noisy operations preserve the equilibrium state. This can be seen in phase space representation, since, for every frequency sector, the covariance matrix of $\tau_M(\bm{\delta})\otimes\tau_{M_E}(\bm{\delta})$ is proportional to the identity (see Eq. \eqref{eq:unifCM}), and the symplectic matrix of $\hat{U}^{(M+M_E)}_{\bm{O}}$ is orthogonal. Clearly, $GN\subseteq UPG$, but also here we do not know whether this inclusion is strict. 

We introduce a quantifier for the resource of non-uniformity as follows: 
\begin{definition}
 A function $\P$, mapping density operators to real numbers, is a \emph{non-uniformity monotone} if
\begin{description}
 \item[(P1)] $\P$ is non-negative and vanishes for the uniform state (see Theorem \ref{theo:unif}). 
 \item[(P2)] $\P$ does not increase under the chosen set of free operations, for instance UP (see Def. \ref{def:UPO}), i.e.
 \begin{equation}
  \P(\Lambda_{UP}(\rho))\leq \P(\rho)\qquad\forall\:\Lambda_{UP}.
 \end{equation}
\end{description}
We define \emph{Gaussian non-uniformity monotones} as those functions $\P^G$ satisfying (P1) and (P2) for uniformity-preserving Gaussian operations (see Def. \ref{def:UPO}).
\end{definition}

In analogy with coherence, we introduce the \emph{relative entropy of non-uniformity}:
\begin{equation}
\label{eq:ps}
\P_{rel}(\rho):= S\left(\rho\|\tau_M(\bm{\delta})\right).
\end{equation}
This function clearly satisfies (P1). The property (P2) follows from the contractivity of the relative entropy, 
 \begin{equation}
  \begin{split}
   \P_{rel}(\rho)&= S\left(\rho\,\|\,\tau_M(\bm{\delta})\right)\geq S(\Lambda_{UP}[\rho]\,\|\,\Lambda_{UP}[\tau_M(\bm{\delta})])= S(\Lambda_{UP}[\rho]\,\|\,\tau_M(\bm{\delta})) =\P_{rel}(\Lambda_{UP}[\rho]).
  \end{split}
 \end{equation}
 
Restricting ourselves to Gaussian states and operations, we find results for the relative entropy of Gaussian non-uniformity $\P_{rel}^G(\rho)$ (that is the relative entropy of non-uniformity for Gaussian states).
\begin{theorem}
\label{theo:Ghier0}
The relative entropy of Gaussian non-uniformity (see Eq. \eqref{eq:ps}) of a Gaussian state $\rho$ is equal to its maximal coherence (see Eq. \eqref{cohmax}):
\begin{equation}
 \P_{rel}^G(\rho)=\C_{rel;\:max}^G(\rho).
\end{equation} 
\end{theorem}
This result follows from Theorem \ref{theo:dist}: 
 \begin{equation}
  \C^G_{rel;\:max}(\rho)=\sum_{\omega=\omega_1}^{\omega_{M_f}}\, S\left(\rho_\omega\,\Big\|\,\tau_{M_s}\left(\frac{N_\omega}{M_s},\dots, \frac{N_\omega}{M_s}\right)\right)= S\left(\rho\|\tau_M(\bm{\delta})\right)=\P_{rel}(\rho),
 \end{equation}
and establishes a strong connection between coherence and non-uniformity for Gaussian systems, in analogy to DV systems \cite{SKWGB18}.

We conclude this section by noticing two additional properties of the relative entropy of Gaussian non-uniformity, which can be found by employing Theorem \ref{theo:Ghier0}:
\begin{itemize}
 \item Pure Gaussian states $\proj{\psi_G}$ maximise the relative entropy of Gaussian non-uniformity among the states with given average occupation number $N_\omega$ and total number of spatial modes $M_s$:
 \begin{equation}
 \begin{split}
  \P_{rel}^G(\proj{\psi_G})&= S(\proj{\psi_G}\|\,\tau_M(\bm{\delta}))\\
  &=\sum_{\omega=\omega_1}^{\omega_{M_f}}\, \left[\left(N_\omega+M_s\right)\,\log\left(\frac{N_\omega}{M_s}+1\right)-N_\omega\log\frac{N_\omega}{M_s} \right],
 \end{split}
\end{equation}
which follows from Eq. \eqref{eq:CSrho}.
\item The relative entropy of Gaussian non-uniformity is invariant under passive Gaussian unitaries, i.e. 
\begin{equation}
\P_{rel}^G(\rho)=\P_{rel}^G(\hat{U}_{\bm O}\,\rho\,\hat{U}_{\bm O}^{\dagger}) 
\end{equation}
This property follows by noticing that the maximal Gaussian coherence cannot be increased via passive Gaussian unitaries (see Eq. \eqref{maxcoh}).
\end{itemize}

\section{Hierarchy of quantum resources in CV systems}
\label{sec:conn}
The relative entropy also quantifies, for CV systems, multipartite entanglement \cite{ESP02} and symmetric quantum discord \cite{modi2010unified, BLA19}: 
\begin{equation}
\label{eq:relentdisc}
\D_{rel}(\rho)=\inf_{\sigma\in\mathcal{Z}} S(\rho \|\sigma) ,
\end{equation}
\begin{equation}
\label{eq:relentent}
\mathcal{E}_{rel}(\rho)=\inf_{\sigma\in\mathcal{S}} S(\rho \|\sigma),
 \end{equation}
where $\mathcal{Z}$ and $\mathcal{S}$ denote the sets of zero-discord and separable states, respectively. The former contains all mixtures of pure, locally orthonormal projectors \cite{modi2010unified, modi2012discord}, while the latter contains all convex combinations of arbitrary product states \cite{braunstein2005quantum}, i.e.
\begin{align}
&\mathcal{Z} \ni \rho =  \sum_{\bm{m}} p_{\bm{m}}\, \proj{\psi_{m_1}}\otimes \proj{\psi_{m_2}}\otimes\dots\otimes\proj{\psi_{m_M}}, \label{eq:zerodis}\\
&\mathcal{S} \ni \rho =  \sum_{\bm{m}} p_{\bm{m}}\, \rho_{m_1}\otimes \rho_{m_2}\otimes\dots\otimes\rho_{m_M} , \label{eq:zeroent}
\end{align}
where $\braket{\psi_{m_k}|\psi_{l_k}}=\delta_{ml}$ for all $k=1,\dots,M$, $\bm{m}:= m_1 m_2 \dots m_M$, $p_{\bm{m}}\geq 0$ and $\sum_{\bm{m}} p_{m_1 m_2 \dots m_M}=1$. Note that $\mathcal{Z}$ is non-convex, since the convex combination of two sets of orthonormal projectors is not, in general, orthonormal. 

Using the relative entropy, we can extend the ordering of resources for discrete-variable to continuous-variable systems:
\begin{equation}
\label{eq:nGhier}
 \P_{rel}(\rho)\geq\C_{rel}(\rho)\geq\mathcal{D}_{rel}(\rho)\geq\mathcal{E}_{rel}(\rho).
\end{equation}

\begin{figure}[htb]
  \centering
   \includegraphics[scale=0.15]{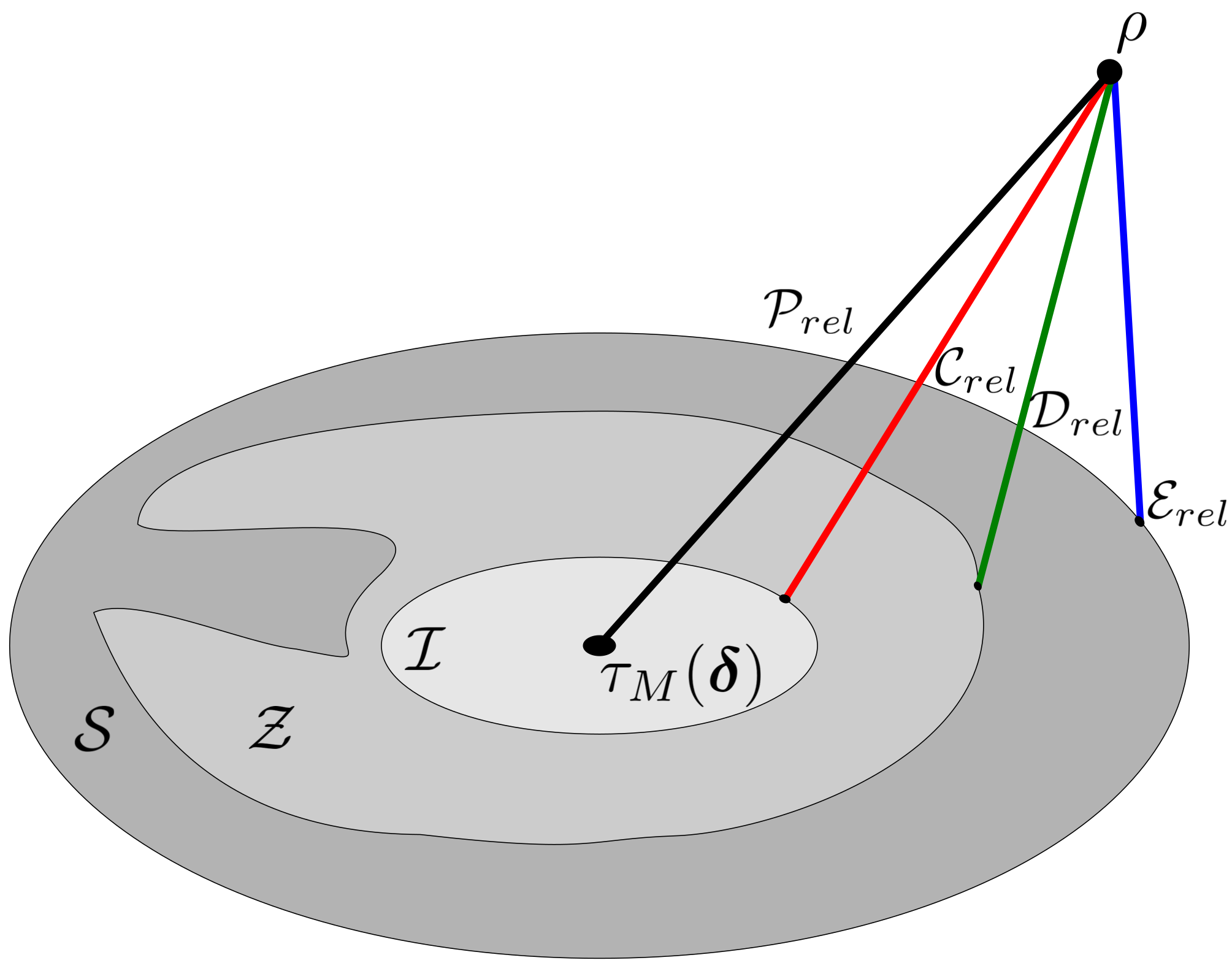}
  \caption{Graphical depiction of the relative entropy of non-uniformity $\P_{rel}$ (black line), coherence $\C_{rel}$ (red line), symmetric quantum discord $\mathcal{D}_{rel}$ (green line) and entanglement $\mathcal{E}_{rel}$ (blue line) for a quantum state $\rho$. The uniform state $\tau_M(\bm{\delta})$ is an element of the incoherent set $\I$, which is a convex subset of the zero-discord set $\mathcal{Z}$, which in turn is a non-convex subset of the separable set $\mathcal{S}$.}
    \label{fig:nGhier}
\end{figure}

This relation directly follows by noting that $\tau_M(\bm{\delta})\in\I\subset\mathcal{Z}\subset\mathcal{S}$, and holds for all quantum states $\rho$ (see Fig. \ref{fig:nGhier}).

Let us now consider the Gaussian case. Let $\mathcal{D}_{rel}^G$ and $\mathcal{E}_{rel}^G$ be the relative entropies of Gaussian discord and entanglement, respectively. They are obtained with Eq. \eqref{eq:relentdisc} and \eqref{eq:relentent} by performing the minimization over the Gaussian subsets $\mathcal{Z}_G$ and $\mathcal{S}_G$ of $\mathcal{Z}$ and $\mathcal{S}$, respectively. While $\mathcal{S}_G$ is defined analogously to Eq. \eqref{eq:zeroent}, by taking Gaussian states, $\mathcal{Z}_G$ is formed by product Gaussian states \cite{AD10, BLA19}, i.e.
\begin{align}
&\mathcal{Z}_G \ni \rho = \rho_{m_1}\otimes \rho_{m_2}\otimes\dots\otimes\rho_{m_M}, \label{eq:Gauszerodis}\\
&\mathcal{S}_G \ni \rho =  \sum_{\bm{m}} p_{\bm{m}}\, \rho_{m_1}\otimes \rho_{m_2}\otimes\dots\otimes\rho_{m_M} , \label{eq:Gausep}
\end{align}
where $\rho_{m_1},\rho_{m_2} \dots,\rho_{m_M}$ and $\rho$ are Gaussian states, and with $\bm{m}:= m_1 m_2 \dots m_M$, $p_{\bm{m}}\geq 0$ and $\sum_{\bm{m}} p_{\bm{m}}=1$ in Eq. \eqref{eq:Gausep}. Since $\tau_M(\bm{\delta})\in\I_G\subset\mathcal{Z}_G\subset\mathcal{S}_G$, Eq. \eqref{eq:nGhier} holds also in this case.

We have discussed in Sec. \ref{sec:MCMS} how passive unitaries can generate coherence. It is well established that they can also generate entanglement \cite{PhysRevLett90047904} and discord \cite{GP10}.

Let us introduce 
\begin{align}
  &\mathcal{D}_{rel;\:max}^G(\rho):=\sup_{\hat{U}_{\bm O}}\,\mathcal{D}_{rel}^G(\hat{U}_{\bm O}\,\rho\,\hat{U}_{\bm O}^{\dagger}), \label{eq:maxdisc}\\
  &\mathcal{E}_{rel;\:max}^G(\rho):=\sup_{\hat{U}_{\bm O}}\,\mathcal{E}_{rel}^G(\hat{U}_{\bm O}\,\rho\,\hat{U}_{\bm O}^{\dagger}). \label{eq:maxent}
\end{align}
\begin{figure}[htb]
  \centering
   \includegraphics[scale=0.2]{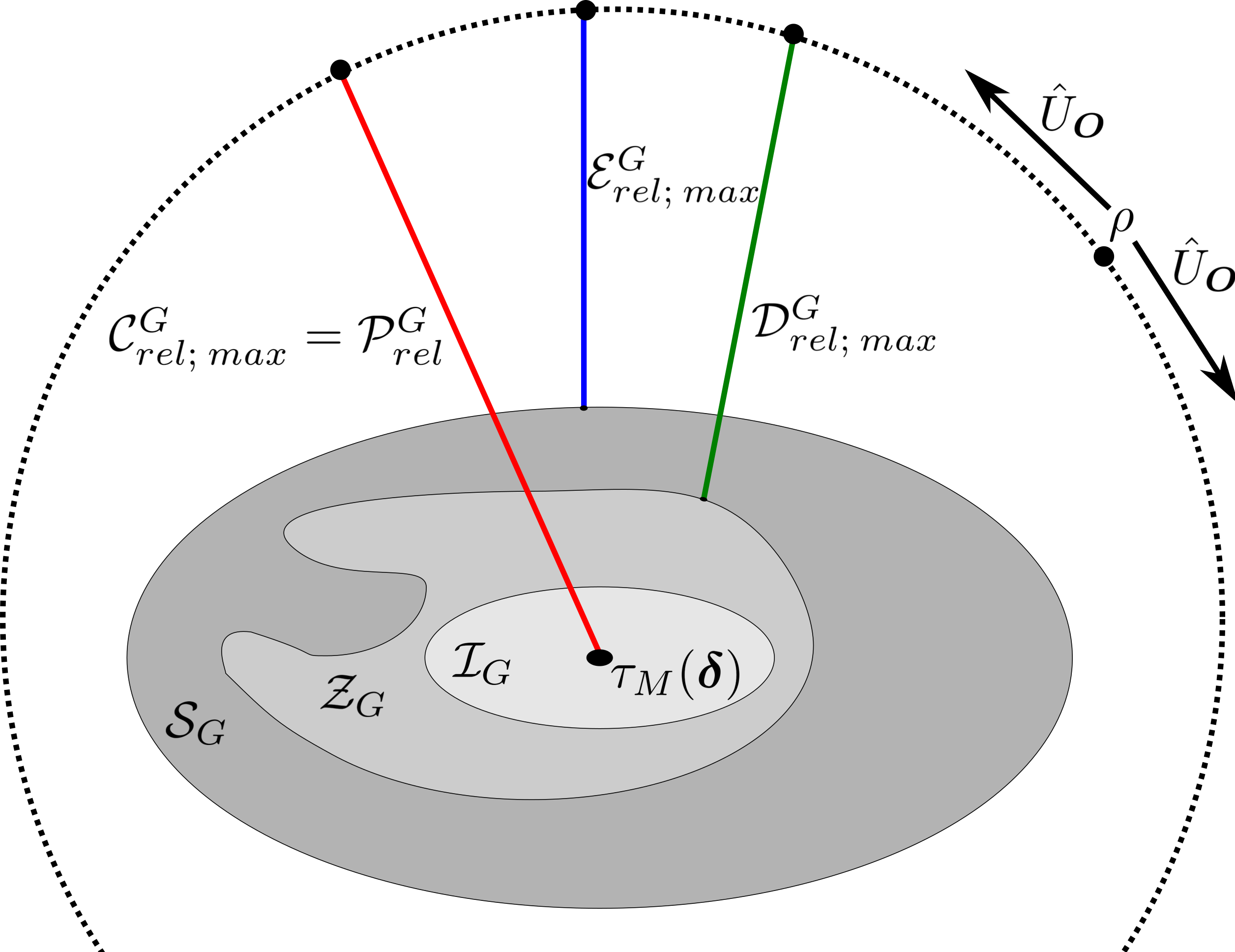}
  \caption{Graphical depiction of Eq. \eqref{eq:Ghier}. The dotted circle represents all the states that can be obtained from $\rho$ via passive unitaries $\hat{U}_{\bm O}$. The red line, connecting the uniform state $\tau_M(\bm{\delta})$ to the MCMGS, is the maximal Gaussian coherence $\C_{rel;\:max}^G(\rho)=\P_{rel}^G(\rho)$. The green and blue lines are the maximal Gaussian discord $\mathcal{D}_{rel;\:max}^G$ and entanglement $\mathcal{E}_{rel;\:max}^G$, respectively. The uniform state $\tau_M(\bm{\delta})$ is an element of the Gaussian incoherent set $\I_G$, which is a convex subset of the Gaussian zero-discord set $\mathcal{Z}_G$, which in turn is a non-convex subset of the Gaussian separable set $\mathcal{S}_G$.}
    \label{fig:Ghier}
\end{figure}

We prove in Appendix \ref{proof:Ghier} the following hierarchy between the mentioned CV resources (see Fig. \ref{fig:Ghier}):
\begin{theorem}
 The relative entropy of Gaussian non-uniformity $\P_{rel}^G$ (Eq. \eqref{eq:ps}) of any Gaussian state $\rho$ is equal to the maximal relative entropy of Gaussian coherence $\C^G_{rel;\:max}$ (Eq. \eqref{cohmax}), and this quantity upperbounds the maximal relative entropies of Gaussian symmetric discord $\mathcal{D}^G_{rel;\:max}$ (Eq. \eqref{eq:maxdisc}) and Gaussian entanglement $\mathcal{E}^G_{rel;\:max}$ (Eq. \eqref{eq:maxent}):
\begin{equation}
\label{eq:Ghier}
  \P_{rel}^G(\rho)=\C^G_{rel;\:max}(\rho)\geq\mathcal{D}^G_{rel;\:max}(\rho)\geq\mathcal{E}^G_{rel;\:max}(\rho)  .
\end{equation}
\end{theorem}
Here we discussed the action of energy-preserving unitaries, in particular passive Gaussian unitaries. An energy non-preserving unitary can, in principle, increase the energy indefinetely and create infinite resources. However, for a fixed finite energy, the ordering of Eq. \eqref{eq:nGhier} is preserved, because the ordering of the sets remains. In the Gaussian scenario, we conjecture that active unitaries exist that keep the hierarchic ordering in Eq. \eqref{eq:Ghier}. The verification of this claim is an interesting open question.

\section{Conclusions}
\label{sec:conc}
In this manuscript, we extended a hierarchy of dicrete-variable quantum resources to continuous-variable systems, under the condition of fixed energy. Considering Gaussian states and operations and using quantifiers based on the relative entropy, we found that the Gaussian non-uniformity is equal to the maximal Gaussian coherence, and we provided an analytical expression for this quantity. This means that, if we quantify the resources with the relative entropy, any amount of Gaussian non-uniformity can be converted into Gaussian coherence by means of a suitable energy-preserving Gaussian unitary. To quantify the non-uniformity, we designed a resource theory by identifying purity at fixed energy as resource. We also considered generic (non-Gaussian) states and found that the non-uniformity always upper-bounds the coherence. Finally, we showed that, for Gaussian states the non-uniformity and the maximal coherence provide upper bounds on the maximal symmetric quantum discord and the maximal entanglement. Our results advance the field of continuous-variable resource theories and establish a further connection between quantum thermodynamics and quantum information theory. 

Our work leaves some interesting questions open. A possible next step could be to study the hierarchy of resources in the presence of energy-nonpreserving Gaussian unitaries, up to a finite maximum energy. In addition, one should investigate whether the equality of maximal coherence and non-uniformity also holds in the general non-Gaussian case. In order to achieve this a deeper understanding of non-Gaussian resource theories is required.

\section*{Acknowledgements}
DB acknowledges inspiring discussions with participants of the Central-European Workshop on Quantum Optics (CEWQO 2019) in Paderborn, in particular with Christine Silberhorn. This research was partially supported by the German Federal Ministry of Education and Research (BMBF), within the project HQS, and by the EU H2020 QuantERA ERA-NET Cofund in Quantum Technologies, within the project QuICHE.

\appendix

\section{Proof of Theorem \ref{theo:mcmsS}}
\label{proof:mcmsS}
For simplicity of notation, we shall drop in the proof the subscript $\omega$ in $\bar{n}_{\omega; j}$, $\rho_\omega$ and $N_\omega$. 

The theorem can be proven with a constrained optimization. Let $\bm{n}=\set{\bar{n}_{1},\dots,\bar{n}_{M_s}}$ and $\mathcal{L}(\bm{n},\lambda)$ be the Lagrangian function
 \begin{equation}
  \begin{split}
  \mathcal{L}(\bm{n},\lambda):=& \C_{rel}^G(\rho;\bm{n}) -\lambda\left(N-\sum_{j=1}^{M_s}\bar{n}_{j}\right) \\ 
  =&S(\rho)+\sum_{j=1}^{\tilde{M}_s}\left[(\bar{n}_{j}+1) \log(\bar{n}_{j}+1)-\bar{n}_{j}\log\bar{n}_{j}\right]+\lambda\sum_{j=1}^{\tilde{M}_s}\bar{n}_{j}-\lambda N,
  \end{split}
 \end{equation}
where $\tilde{M}_s$ is the number of modes for which $\bar{n}_{j}\neq 0$. Since $S(\rho)$ depends only on the symplectic spectrum (see Eq. \ref{eq:entr}), it holds
 \begin{equation}
  \frac{\partial \mathcal{L}(\bm{n},\lambda)}{\partial \bar{n}_{j}}= 
  \begin{cases}
  \log\left(\frac{\bar{n}_{j}+1}{\bar{n}_{j}}\right)+\lambda \qquad \text{for $\bar{n}_{j}\neq 0$}\\
  0 \qquad \text{for $\bar{n}_{j}= 0$}
  \end{cases}
\end{equation} 
 The condition $\partial \mathcal{L}/\partial \bar{n}_{j}=0$ for $\bar{n}_{j}\neq 0$ is equivalent to
 
 \begin{equation}
  \lambda=-\log\left(\frac{\bar{n}_{j}+1}{\bar{n}_{j}}\right),\qquad \forall j
 \end{equation}
 
 The above relations are satisfied by any state $\rho^*$ with $\tilde{M}_s$ reduced occupation numbers $\bar{n}_{j}=N/\tilde{M}_s\quad \forall\, j=1,\dots\tilde{M}_s$ and the others equal to zero. Clearly $1\leq \tilde{M}_s \leq M_s$. 
 
 The Gaussian coherence of any $\rho^*$ reads:
 \begin{equation}
\begin{split}
 \C_{rel}^G(\rho^*)&=-S(\rho^*)+(N+\tilde{M}_s)\log\left(\frac{N+\tilde{M}_s}{\tilde{M}_s}\right)-N\log\left(\frac{N}{\tilde{M}_s}\right).
\end{split}
\end{equation}
Taking the derivative of this expression with respect to $\tilde{M}_s$, i.e.
 \begin{equation}
 \frac{d \C_{rel}^G(\rho^*)}{d\tilde{M}_s}=\log\left(\frac{N+\tilde{M}_s}{\tilde{M}_s}\right)>0,
\end{equation}
one can verify that this function is monotonically increasing with $\tilde{M}_s$. 

The minimum is therefore obtained for $\tilde{M}_s=1$, i.e. when $N$ is contained in a single reduced spatial mode, and the maximum is obtained for $\tilde{M}_s=M_s$, i.e. when $N$ is equally distributed.

\section{Proof of Theorem \ref{theo:dist}}
\label{proof:dist}

Let $\hat{U}_{\mathcal{C}}=\bigotimes_{\omega=\omega_1}^{\omega_{M_f}}\,\hat{U}_{\mathcal{C} \omega}$ be the passive unitary that maximises the relative entropy of Gaussian coherence for $\rho$, i.e. 
\begin{equation}
 \rho_{max}=\hat{U}_{\mathcal{C}}\,\rho\,\hat{U}_{\mathcal{C}}^{\dagger}=\bigotimes_{\omega=\omega_1}^{\omega_{M_f}}\,\hat{U}_{\mathcal{C} \omega}\,\rho_\omega\,\hat{U}_{\mathcal{C} \omega}^{\dagger}.
\end{equation}
Let $\tau_M(\bar{\bm n}_{\rho_{max}})$ be the thermal state with the same $\bar{n}_{\omega;j}$ as $\rho_{max}$. Using Theorem \ref{theo:mcmsS}, $\tau_M(\bar{\bm n}_{\rho_{max}})$ reads
\begin{equation}
 \tau_M(\bar{\bm n}_{\rho_{max}}):=\bigotimes_{\omega=\omega_1}^{\omega_{M_f}}\, \tau_{M_s}(\bar{\bm n}^\omega_{\rho_{max}}),\qquad
 \bar{\bm n}^\omega_{\rho_{max}}:=\left(\frac{N_\omega}{M_s},\frac{N_\omega}{M_s}, \dots,\frac{N_\omega}{M_s}\right).
\end{equation}
Using Eq. \eqref{eq:CSrho} we get
\begin{equation}
 \begin{split}
  \C_{rel}^G(\rho_{max})=& S\left(\rho_{max}\,\|\,\tau_M(\bar{\bm n}_{\rho_{max}})\right)
  =-S(\rho_{max})+\mathrm{Tr}\left[\rho_{max}\log\, \tau_M(\bar{\bm n}_{\rho_{max}})\right]=\\
  =&-S(\rho)+\sum_{\omega=\omega_1}^{\omega_{M_f}}\,\mathrm{Tr}\left[\hat{U}_{\mathcal{C}\omega}\,\rho_\omega\,\hat{U}_{\mathcal{C}\omega}^{\dagger}\,\log\,\tau_{M_s}(\bar{\bm n}^\omega_{\rho_{max}})\right]\\
  =&-S(\rho)+\sum_{\omega=\omega_1}^{\omega_{M_f}}\,\mathrm{Tr}\left[\rho_\omega\, \log\,\left(\hat{U}_{\mathcal{C}\omega}^{\dagger}\,\tau_{M_s}(\bar{\bm n}^\omega_{\rho_{max}})\,\hat{U}_{\mathcal{C}\omega}\right)\right].
 \end{split}
\end{equation}

Here, we have used the invariance of entropy under unitary operation and the diagonality of $\tau(\bar{\bm n}^\omega_{\rho_{max}})$ in the Fock basis.
We then notice that 
\begin{equation}
 \hat{U}_{\mathcal{C}\omega}^{\dagger}\,\tau_{M_s}\left(\bar{\bm n}^\omega_{\rho_{max}}\right)\,\hat{U}_{\mathcal{C}\omega}=\tau_{M_s}\left(\bar{\bm n}^\omega_{\rho_{max}}\right).
\end{equation}
This can be proven by using the phase space representation, since the covariance matrix of $\tau_{M_s}(\bar{\bm n}^\omega_{\rho_{max}})$ is proportional to the identity (consider equal $\bar{n}_m$ in Eq. \eqref{eq:therCM}), and passive Gaussian unitaries are associated to symplectic orthogonal matrices, by Def. \ref{def:pasGau}. It follows that
\begin{equation}
 \begin{split}
  \C_{rel}^G(\rho_{max})=& -S(\rho)+ \sum_{\omega=\omega_1}^{\omega_{M_f}}\,\mathrm{Tr}\left(\rho_\omega\,\log\,\tau_{M_s}(\bar{\bm n}^\omega_{\rho_{max}})\right)\\ 
  =& -\sum_{\omega=\omega_1}^{\omega_{M_f}}\,S(\rho_\omega)+ \sum_{\omega=\omega_1}^{\omega_{M_f}}\,\mathrm{Tr}\left(\rho_\omega\,\log\,\tau_{M_s}(\bar{\bm n}^\omega_{\rho_{max}})\right) \\
  =&\sum_{\omega=\omega_1}^{\omega_{M_f}}\, S\left(\rho_\omega\,\Big\|\,\tau_{M_s}\left(\frac{N_\omega}{M_s},\dots, \frac{N_\omega}{M_s}\right)\right).
 \end{split}
\end{equation}
\section{Maximal Gaussian coherence for specific states}
\label{proof:examples}
In this section, we will consider modes with the same frequency and drop the subscript $\omega$. By Theorem \ref{theo:mcmsS}, we can search for a passive unitary that equally distributes the average occupation number $N$ of a Gaussian state $\rho$ among its modes: this unitary maximises the relative entropy of coherence of $\rho$. 

As a first case, let us consider a generic two-mode Gaussian state $\rho$, with mode operators $\a_1$ and $\a_2$. We now apply a $50:50$ beam splitter of phase $\phi$ (to be specified later):

\begin{equation}
 \begin{split}
 \a_1 &\mapsto \hat{b}_1=\frac{1}{\sqrt{2}}\,\a_1+ \frac{e^{\mathrm{i}\phi}}{\sqrt{2}}\,\a_2\\
 \a_2 & \mapsto \hat{b}_2=\frac{1}{\sqrt{2}}\,\a_2-\frac{e^{-\mathrm{i}\phi}}{\sqrt{2}}\,\a_1  
 \end{split}
\end{equation}

Then we have
 
\begin{equation}
 \begin{split}
 \braket{\hat{b}^\dagger_1\hat{b}_1} &=\frac{1}{2}\,\braket{\ad_1\a_1} + \frac{1}{2}\,\braket{\ad_2\a_2} +\frac{e^{\mathrm{i}\phi}}{2}\braket{\ad_1\a_2} + \frac{e^{-\mathrm{i}\phi}}{2}\braket{\a_1\ad_2}\\ 
 &= \frac{N}{2}+ \mathrm{Re} \left( e^{-\mathrm{i}\phi}\braket{\a_1\ad_2}\right) \\
 \braket{\hat{b}^\dagger_2\hat{b}_2}&=\frac{1}{2}\,\braket{\ad_2\a_2}+\frac{1}{2}\,\braket{\ad_1\a_1}-\frac{e^{-\mathrm{i}\phi}}{2}\braket{\ad_2\a_1} - \frac{e^{\mathrm{i}\phi}}{2}\braket{\a_2\ad_1} \\
 &= \frac{N}{2}- \mathrm{Re} \left( e^{-\mathrm{i}\phi}\braket{\a_1\ad_2}\right) .
 \end{split}
\end{equation}
With $\braket{\a_1\ad_2}=|\braket{\a_1\ad_2}|e^{\mathrm{i}\theta_{12}}$, we can choose $\phi=\frac{\pi}{2}-\theta_{12}$, leading to $\braket{\hat{b}^\dagger_1\hat{b}_1}=\braket{\hat{b}^\dagger_2\hat{b}_2}=N/M_s$ (in this case $M_s=2$), thus maximising the coherence. 

Let us now consider a generic $M_s$-mode product state $\rho=\varrho_1\otimes\dots\otimes\varrho_{M_s}$ with $\bm{d}=0$. We prove that the \emph{quantum Fourier transform} (QFT)
\begin{equation}
\a_{j}\mapsto \hat{b}_{j}:=\sum_{k=1}^{M_s} U_{jk}\,\a_{j k}=\frac{1}{\sqrt{M_s}}\,\sum_{k=1}^{M_s} e^{\frac{2\pi\,\mathrm{i}}{M_s}\,(j-1)(k-1)}\,\a_{j}
\end{equation}
is the passive Gaussian unitary that maximises the coherence. After the action of the QFT, the occupation number for the mode $j$ reads 
\begin{equation}
 \braket{\hat{b}_j^\dagger\, \hat{b}_j}=\frac{1}{M_s}\sum_{k,k'=1}^{M_s} \, e^{\frac{2\pi\,\mathrm{i}}{M_s}\,(j-1)(k'-k)}\,\braket{\a^\dagger_k\,\a_{k'}}.
\end{equation}
We separate the sum into two parts, with $k=k'$ and $k\neq k'$:
\begin{equation}
\begin{split}
 \braket{\hat{b}_j^\dagger\, \hat{b}_j}=&\frac{1}{M_s}\sum_{k=1}^{M_s} \, \braket{\a^\dagger_k\,\a_{k}}+\frac{1}{M_s}\sum_{k\neq k'} \, e^{\frac{2\pi\,\mathrm{i}}{M_s}\,(j-1)(k'-k)}\,\braket{\a^\dagger_k\,\a_{k'}} \\
 =&\frac{N}{M_s}+\frac{1}{M_s}\sum_{k\neq k'} \, e^{\frac{2\pi\,\mathrm{i}}{M_s}\,(j-1)(k'-k)}\,\braket{\a^\dagger_k\,\a_{k'}}.
 \end{split}
\end{equation}

Since $\rho$ is a product state and $\bm{d}=0$, we conclude the proof by noticing
\begin{equation}
 \braket{\a^\dagger_k\,\a_{k'}}=\braket{\a^\dagger_k}\braket{\a_{k'}}=0.
\end{equation}

Remarkably, this transformation is an extension of the DV unitary to CV. The unitary that maximises the coherence of an arbitrary DV state $\rho$ for any MIO monotone reads \cite{SKWGB18} 
\begin{equation}
 \hat{U}_{max}^{DV}=\frac{1}{\sqrt{d}}\,\sum_{n=1}^d\,\sum_{k=1}^{d}\, e^{\frac{2\pi\,\mathrm{i}}{M_s}\,(n-1)(k-1)}\ketbra{k}{\rho_n},
\end{equation}
where $d$ is the dimension of $\rho$, $\ket{\rho_n}$ are the eigenstates of $\rho$, and $\ket{k}$ are the elements of the incoherent basis. Notice, however, that the CV result only holds for product states with $\bm{d}=0$. 
\section{Proof of Theorem \ref{theo:unif}}
\label{proof:unif}
This proof is similar to that of Theorem \ref{theo:mcmsS}, and also here we drop the subscript $\omega$ in $\bar{n}_{\omega; j}$ and $N_\omega$.

The covariance matrix of a Gaussian thermal state (see Eq. \eqref{eq:therCM}) is diagonal and coincides with the diagonal matrix $\bm{D}$ in Williamson's theorem (see Eq. \eqref{eq:WT}). From Eq. \eqref{eq:entr}, it follows that the von-Neumann entropy of $\tau_{M_s}(\bar{\bm n})$ reads
\begin{equation}
S\left(\tau_{M_s}(\bar{\bm n})\right)=\sum_{j=1}^{\tilde{M}_s}\,\left[(\bar{n}_j+1)\,\log(\bar{n}_j+1)-\bar{n}_j\,\log\bar{n}_j\right],
\end{equation}
where $\bm{n}=(\bar{n}_{1},\dots,\bar{n}_{M_s})$ and $\tilde{M}_s$ is the number of modes for which $\bar{n}_{j}\neq 0$. Let $\mathcal{L}(\bm{n},\lambda)$ be the Lagrangian function
 \begin{equation}
  \begin{split}
  \mathcal{L}(\bm{n},\lambda):=& S\left(\tau_{M_s}(\bar{\bm n})\right) -\lambda\left(N-\sum_{j=1}^{\tilde{M}_s}\bar{n}_{j}\right) \\ 
  =&\sum_{j=1}^{\tilde{M}_s}\left[(\bar{n}_{j}+1) \log(\bar{n}_{j}+1)-\bar{n}_{j}\log\bar{n}_{j}\right]+\lambda\sum_{j=1}^{\tilde{M}_s}\bar{n}_{j}-\lambda N,
  \end{split}
 \end{equation}
It holds 
 \begin{equation}
  \frac{\partial \mathcal{L}(\bm{n},\lambda)}{\partial \bar{n}_{j}}= 
  \begin{cases}
  \log\left(\frac{\bar{n}_{j}+1}{\bar{n}_{j}}\right)+\lambda \qquad \text{for $\bar{n}_{j}\neq 0$}\\
  0 \qquad \text{for $\bar{n}_{j}= 0$}
  \end{cases}
\end{equation} 
 The condition $\partial \mathcal{L}/\partial \bar{n}_{j}=0$ for $\bar{n}_{j}\neq 0$ is equivalent to
 
 \begin{equation}
  \lambda=-\log\left(\frac{\bar{n}_{j}+1}{\bar{n}_{j}}\right),\qquad \forall\: j.
 \end{equation}
 
 The above relations are satisfied by any state $\tau_{M_s}(\bar{\bm n}^*)$ with $\tilde{M}_s$ reduced occupation numbers $\bar{n}_{j}=N/\tilde{M}_s\quad \forall\, j=1,\dots\tilde{M}_s$ and the others equal to zero. Clearly $1\leq \tilde{M}_s \leq M_s$. 
 
 The entropy of any $\tau_{M_s}(\bar{\bm n}^*)$ reads:
 \begin{equation}
\begin{split}
 S\left(\tau_{M_s}(\bar{\bm n}^*)\right)&=(N+\tilde{M}_s)\log\left(\frac{N+\tilde{M}_s}{\tilde{M}_s}\right)-N\log\left(\frac{N}{\tilde{M}_s}\right).
\end{split}
\end{equation}
Taking the derivative of this expression with respect to $\tilde{M}_s$
 \begin{equation}
 \frac{d S\left(\tau_{M_s}(\bar{\bm n}^*)\right)}{d\tilde{M}_s}=\log\left(\frac{N+\tilde{M}_s}{\tilde{M}_s}\right)>0,
\end{equation}
one can verify that this function is monotonically increasing with $\tilde{M}_s$. 

The minimum is therefore obtained for $\tilde{M}_s=1$, i.e. when $N$ is contained in a single reduced spatial mode, and the maximum is obtained for $\tilde{M}_s=M_s$, i.e. when $N$ is equally distributed.

\section{Proof of Eq. (\ref{eq:Ghier})}
\label{proof:Ghier}

Let $\hat{U}_\mathcal{E}$ be the passive Gaussian unitary that achieves $\mathcal{E}^G_{rel;\:max}(\rho)$ in Eq. \eqref{eq:maxent}. Then we have
\begin{equation}
\begin{split}
 \mathcal{E}^G_{rel;\:max}(\rho)&=\mathcal{E}_{rel}^G\left(\hat{U}_\mathcal{E}\,\rho\,\hat{U}_\mathcal{E}^{\dagger}\right)\leq\mathcal{D}_{rel}^G\left(\hat{U}_\mathcal{E}\,\rho\,\hat{U}_{\mathcal{E}}^{\dagger}\right)\\
 &\leq\sup_{\hat{U}_{\bm O}}\mathcal{D}_{rel}^G\left(\hat{U}_{\bm O}\,\rho\,\hat{U}_{\bm O}^{\dagger}\right)=\mathcal{D}^G_{rel;\:max}(\rho).
\end{split}
 \end{equation}

Similarly, let $\hat{U}_\mathcal{D}$ be the Gaussian unitary that achieves $\mathcal{D}^G_{rel;\:max}(\rho)$ in Eq. \eqref{eq:maxdisc}. Then 
\begin{equation}
\begin{split}
 \mathcal{D}^G_{rel;\:max}(\rho)&=\mathcal{D}_{rel}^G\left(\hat{U}_\mathcal{D}\,\rho\,\hat{U}_{\mathcal{D}}^{\dagger}\right)\leq\mathcal{C}_{rel}^G\left(\hat{U}_\mathcal{D}\,\rho\,\hat{U}_{\mathcal{D}}^{\dagger}\right)\\
 &\leq\sup_{\hat{U}_{\bm O}}\mathcal{C}_{rel}^G\left(\hat{U}_{\bm O}\,\rho\,\hat{U}_{\bm O}^{\dagger}\right)=\mathcal{C}^G_{rel;\:max}(\rho).
 \end{split}
\end{equation}

Finally, using $\C_{rel}^G(\rho_{max})=\P_{rel}^G(\rho)$ from Theorem \ref{theo:Ghier0}, we obtain the desired result.

\printbibliography
\end{document}